\documentstyle[preprint,aps,epsfig]{revtex}
\begin{document}

\title{ Low-dimensional dynamical system model for
observed coherent structures in ocean satellite data}
\author{Crist\'obal L\'{o}pez \footnote[1]{email: clopez@imedea.uib.es}
and Emilio Hern\'andez-Garc\'\i a 
\footnote[2]{email: emilio@imedea.uib.es}}

\address{Instituto Mediterr\'aneo de Estudios Avanzados
IMEDEA (CSIC-UIB),\\ Campus de la Universidad de las Islas
Baleares, E-07071 Palma de Mallorca, Spain.}

\date{\today}
\maketitle

\begin{abstract}
The dynamics of coherent structures present in {\it real-world}
environmental data is analyzed. The method developed in this Paper
combines the power of the Proper Orthogonal Decomposition (POD)
technique to identify these coherent structures in experimental
data sets, and its optimality in providing Galerkin basis for
projecting and reducing complex dynamical models. The POD basis
used is the one obtained from the experimental data. We apply the
procedure to analyze coherent structures 
in an oceanic setting, the ones
arising from
instabilities of the Algerian current, in the western
Mediterranean Sea. Data are from satellite altimetry providing Sea
Surface Height, and the model is a two-layer quasigeostrophic
system. A four-dimensional dynamical system is obtained that
correctly describe the observed coherent structures (moving
eddies). Finally, a bifurcation analysis is performed on the
reduced model.

\end{abstract}

\section{Introduction}

 In the last decades the study of turbulent or extended chaotic systems has
enjoyed important advances. Two of them are, first, the
recognition of the existence and high relevance of coherent
structures (defined as strongly {\sl persistent} spatiotemporal
structures in the dynamical evolution of the system) in weakly and
even strongly chaotic systems, and, second, the borrowing of
mathematical methods coming from the studies of nonlinear
dynamical systems (see \cite{HLB} and references therein).

In both subjects, the introduction of the statistical technique
known as the Proper Orthogonal Decomposition (POD, also known
under a variety of other names, such as Karhunen-Lo\`{e}ve
decomposition, method of Empirical Orthogonal Eigenfunctions,
etc.) has played an important r\^{o}le. It was introduced in the
context of turbulence by Lumley \cite{lu} and has revealed 
itself as an
efficient technique for finding, describing and modeling coherent
structures in turbulent fluids or extended chaotic systems. The
purpose of POD is to separate a given data set into orthogonal
spatial and temporal modes which most efficiently absorb the
variability of the data set.

The power of POD has been implemented following two different
paths \cite{HLB}: On the one hand the POD is used as a standard
technique to extract coherent structures from empirical data sets
\cite{Preisen,Palacios96}. Contrasting to Fourier Decomposition,
the eigenfunctions obtained from the POD may display spatial
localization, and thus provide a more efficient way to represent
coherent structures. The use of empirical information can be
pushed further and methodologies from dynamical systems theory and
other fields have been used in the POD framework, to provide
useful algorithms for control \cite{Qin94} and prediction
\cite{PRLprediction,GRLprediction}. On the other hand, the POD
eigenfunctions provide a set of basis functions which is optimum
(at least in a well defined linear sense) for obtaining
low-dimensional ordinary differential equation (ODE)
approximations starting from models based on partial differential
equations (PDEs). The approximation is performed by obtaining long
runs of the PDEs, performing the POD onto this synthetic data set,
and using the Galerkin method to project the PDE model into the so
obtained POD eigenfunctions
\cite{Sirovich87,Sirovich89,Rodriguez90,Deane91,Sahan97}.

These two potentialities, i.e. the ability to extract empirical
information from experimental data, and the efficiency in building
low-dimensional projections from models, are not frequently used
together in the literature. A remarkable exception is the use of
empirical eigenfunctions obtained from the POD of experimental
data from a turbulent boundary layer to build a low-dimensional
approximation to the Navier-Stokes equations
\cite{Aubry88,Aubry89}. We believe that, in some circumstances,
the projection of theoretical models into experimentally obtained
empirical functions could improve both the model and the data.
This will occur in situations such as in the modeling of natural
phenomena (ocean or atmospheric dynamics, for example) where even
very complex models may be not accurate enough, and data are
unvoidably noisy and difficult to calibrate. Projecting the model
onto the experimental eigenfunctions will force it to stay into
the `right' subspace, providing a kind of data assimilation
\cite{assimilation} that may compensate the loss of details
inherent to low-dimensional projections. On the other hand, the
truncation involved in the POD method implies a kind of filtering
providing noise reduction to the data set.

Our aim in this Paper is to explore the synergy between
experimental observation and low-dimensional reduction via POD, in
the complex setting of environmental fluid dynamics. In
particular, a model for coherent structures arising from
instabilities of the Algerian current in the Mediterranean Sea
will be set up and analyzed.

In the ocean dynamics context, the kind of spatiotemporal data
sets we need for our purposes can only be obtained from satellite
observations. The recent availability of satellite data of the sea
surface is allowing a deeper understanding of the Ocean.
Satellites continuously measure sea temperature, sea level,
chlorophyll concentration, etc., which increase our knowledge of
ocean currents, mean sea level changes, tides, or plankton
dynamics, to name a few. In particular, in the last decade the ERS
and the TOPEX/POSEIDON (T/P) satellite missions have provided the
scientific community with high-accuracy altimetry data, which
determines the sea surface level, this is, the height of the sea
surface over a reference level on Earth. Among other relevant
scientific applications, these type of data are specially useful
for a better understanding of the dynamics of mesoscale phenomena
in the Ocean. Mesoscale refers to typical spatial scales of $30$
to $300$ kilometers and time scales of less than one year, and it
is associated with movements of oceanic currents and short-time
flow variations, and also with the formation and propagation of
ocean eddies. These eddies are generated by interactions with the
oceanic topography and/or mean flow instabilities and their
importance is enormous: for example they play a fundamental role
in the heat transport from low to high latitudes.

Some details of the results we present here are determined by the
peculiarities of the data set we are going to use. In particular
we mainly focus in the motion of a vortex which is present in the
data, but is only described by subdominant eigenfunctions in the
POD results. This lead us to a model for this coherent structure,
but we do not try to model the full dynamics of whole data set. We
expect however that our general methodology will be useful in
other problems in which real-world noisy observations and complex
but imperfect PDE models are available. Generalizations of the POD
method which take into account in a more consistent way dynamic
constraints have been developed \cite{Hasselmann,Uhl,Kwasniok97}
and even applied to geophysical contexts
\cite{Hasselmann,Achatz,Kwasniok96}. 
We will use however, and just for simplicity, the standard
formulation of the POD technique.

The Paper is organized as follows: in the next section, we
describe the data set. In section III, these data are analyzed
with the Proper Orthogonal Decomposition (POD). The study of the
most relevant of the eigenfunctions allows the identification of a
coherent structure, that is, a moving vortex or eddy. Then, in the
following section, the associated temporal modes of the POD
eigenfunctions defining the eddy are projected over a hydrodynamic
model which, finally, provides a deterministic dynamical system
depending on the parameters of the model. In section V, the
reconstruction of the moving vortex from the dynamical system
model is performed. Next, in section VI the bifurcation analysis
of the dynamical system is shown. Section VII concludes this work.

\section{Satellite altimetry data}
\label{data}

 We analyze altimetry data
from the T/P and ERS-1 satellite missions \cite{AVISO}.
Altimetry data of the Ocean provide the Sea Surface Height (SSH)
over a reference substrate. The data obtained in both missions
have been merged (to obtain a better spatiotemporal resolution) on
a common time period, from October 1992 to December 1993 and,
finally, $44$ maps taken every $10$ days on a $0.2^{o}$ regular
grid are obtained for the Western Mediterranean Sea \cite{Ay}. We
restrict our analysis to the area known as the Algerian Current,
localized between $0-15^{o}E$ and $ 35-40^{o}N$, where a strong
mesoscale activity is observed \cite{Bou1}. A mean flow moving
eastwards and parallel to the coast of Algeria is the main feature
in this area. It undergoes instabilities that shed vortices into
the western Mediterranean basin, greatly influencing the physical
and biological processes in this area of the Sea \cite{tintore}.
In Figure 1 a) we show in a small box of an image of the Mediterranean
Sea, the area under study. Figure 1 b) shows 
one of the $44$ maps that we are going to analyse.

The SSH fields allow the identification of coherent structures
(structures approximately maintained in the flow over long time
periods) in geophysical flows. In particular, areas of higher
altimetric values may correspond to anticyclonic (clock-wise)
vortices and lower ones may  indicate the existence of cyclonic
(anticlock-wise) vortices. Actually, the data we have used are
referred to a mean level, i.e, we analyze Sea Level Anomalies
(SLA) where the reference height is the temporal mean of the data.
This may give rise to some  minor problems because to obtaining the
SSH data (the one we are going to need in our modeling approach)
is not as simple as adding the mean sea level. This is because of
the different resolution in the data and will be explained in
detail in section IV.

\section{POD analysis of the satellite data}
\label{POD}

As it has already been mentioned, the POD technique is generally
used to analyze experimental or numerical data with the view in
extracting their dominant features, which will typically be
patterns in space and time. On output, it provides a set of
orthogonal functions which are the eigenfunctions of the
covariance matrix of the data. Generally, this set is ordered in
decreasing size of the corresponding eigenvalue, the larger the
eigenvalue meaning the larger percentage of the data variance is
contained in the dynamics of corresponding eigenfunction. Thus, if
$u(x,y,t)$ is our data field ($(x,y)\in A \subset R^2$ is the
spatial point and $t$ is time) to which the temporal average has
been subtracted,
 the POD
basis $\{ \phi_i(x,y), i=1,...,\infty \}$ is obtained after solving
\begin{equation}
\int_A <u(x,y,t)u(x',y',t)>\phi_i(x',y')dx'dy' = \lambda_i
\phi_i(x,y),
\label{eigenproblem}
\end{equation}
\noindent being $<\cdot >=\frac{1}{T}\int_0^T \cdot $ $dt$, i.e.,
the time average, and $\lambda_i$ the corresponding eigenvalues,
which are ordered in decreasing size $\lambda_1 \ge \lambda_{2}
\ge ...\ge 0$. Therefore, we have the modal decomposition
\begin{equation}
u(x,y,t)=\sum_{i=1}^\infty a_i(t) \phi_i(x,y),
\label{expansion}
\end{equation}

\noindent where the $a_i(t)$ are the so-called temporal modes.
In addition, the
following orthogonality conditions are fulfilled
\begin{equation}
\int_A \phi_k(x,y) \phi_l(x,y) dx dy=\delta_{kl},
\end{equation}
\begin{equation}
\left<a_k(t) a_l(t)\right> = \frac{1}{T}\int_0^T a_k(t) a_l(t) dt =
\lambda_k \delta_{kl},
\end{equation}
\noindent where $\delta_{kl}$ is the Kronecker delta.

The optimality of the POD basis functions means that \cite{HLB},
among all linear decompositions with respect to an arbitrary basis
$\{
\Phi_i(x,y) \} $, for a truncation or order $N$,
i.e., $ u^N (x,y,t)=\sum_{i=1}^N a_i (t) \Phi_i (x,y) $, with  $
a_i(t)=\int_A u(x,y,t) \Phi_i (x,y)$, the minimum error, defining
the error as $\epsilon = \left< \int_A (u-u^N)^2 dx dy \right>$,
is obtained when
$\{\Phi_i(x,y) \} $is the POD basis $\{
\phi_i(x,y) \} $.

We now apply the POD analysis to the altimetry satellite data. The
results of this are outlined in the following.
Fig.~\ref{fig:fullspectrum} shows (in linear-log scale) the
fraction of variance ${\lambda_i}/({\sum_{n=1}^{N}\lambda_n})$,
given by each eigenvalue. It is clearly seen that most of the
variance is captured by the first and second eigenvalues. In
Fig.~\ref{fig:anualmodes} we show the temporal mode associated to
the first two eigenvalues, and the power spectrum of the four
dominant ones is plotted in Fig.~\ref{fig:power}. The annual
seasonal periodicity is clearly observed in the two dominant
modes. This and more detailed observations in terms of a complex
version of the POD on the same data set \cite{Bou1} allow us to
interpret the dynamics given by these dominant eigenfuntions as
the seasonal response of the Ocean (heating in summer and cooling
in winter) along the annual cycle.

Therefore, to get a deeper insight into the data (and in
particular in the mesoscale phenomena), we have to study other
eigenfunctions than the first two. In
Fig.~\ref{fig:partialspectrum} we show again the fraction of
variance of the eigenvalues (linear-log plot) but in this case we
have removed the first two eigenvalues. We observe that
eigenvalues $3^{rd}$ and $4^{th}$ are equivalent under the error
bar (for calculating error bars in POD eigenvalues see
\cite{NBCM}). In this case, a linear superposition of them can
represent a moving coherent structure \cite{Preisen}. In
Fig.~\ref{fig:temporalmodes34} we show the temporal modes
associated with eigenvalues $3^{rd}$ and $4^{th}$. This Figure,
and the corresponding power spectra in Fig.~\ref{fig:power}
suggest a weak semiannual periodicity for both modes.

Visualization of the time evolution of the data filtered to keep
just the eigenfuntions  $3^{rd}$ and $4^{th}$, i.e.,
$u_{34}(x,y,t)=a_3(t)
\phi_3 (x,y)+ a_4 (t) \phi_4(x,y)$, suggest a vortex moving northward and
eastwards from the Algerian coast, in agreement with the results
of \cite{Bou1}. The approximate periodicity of the associated
temporal modes indicates that new vortices are shed from the coast
roughly every six months. Describing the dynamics of this coherent
structure found in the flow will be our goal in the remaining of
the Paper.

\section{Model and low-dimensional dynamical system}
\label{low}

The classical way of  using POD to obtain a low-dimensional
dynamical system approximation, which takes into account the most
relevant features of the physical system, consists in truncating
the expansion (\ref{expansion}) to a particular order
\cite{HLB,Sirovich87,Sirovich89,Rodriguez90,Deane91,Sahan97,Aubry88,Aubry89}.
This order is generally chosen to contain most of the percentage
of the variance of the data. Then, the equations governing the
dynamics of the system (PDEs from which the data may have been
generated numerically) are projected over this particular Galerkin
basis and a system of ODEs for the temporal modes $a_i (t)$ can be
obtained. Our approach is somewhat different, first of all, our
data are from satellite observations and we need a specific
mathematical model to describe approximately our data, and second,
our interest focuses in the dynamics associated with the
eigenfuntions $3^{rd}$ and $4^{th}$ which seem to contain the
evolution of the moving mesoscale vortex, and not the more
dominant eigenfunctions $1^{st}$ and $2^{nd}$.

Proceeding with the modeling step, and supported by marine
experimental campaigns \cite{Bou1,Algers}, we assume that the
strong mesoscale activity in the Algerian current is mainly due to
baroclinic instability phenomena. This name refers to the
instabilities grown from the available potential energy associated
with horizontal gradients of density \cite{Ped}. Therefore, we
choose a two-layer quasigeostrophic model as our basic flow
description since this is the minimal model accounting for these
type of instabilities. Layered quasigeostrophic models are widely
used in oceanographical modeling and their main assumption is that
the Ocean behaves as having different layers where density is
constant and, in all the different layers, geostrophic balance is
maintained (i.e. Coriolis and pressure forces nearly equilibrate
via a quasibidimensional flow). Actually, the spatial scales of
the chosen region, and its strong topographic features, lead
however to important deviations from quasigeostrophy. Thus, the
postulated model should be considered at most as a crude
approximation to the real dynamics. It is one of the objectives of
this Paper to show that the the empirical information contained in
the satellite data is incorporated into the model during the
projection procedure, so that the final low-dimensional model
gives a reasonable description of the dynamics.

In the framework of multilayer quasigeostrophic models, every
fluid layer $i$ of density $\rho_i $ and thickness $h_i$ is
described by a stream function $\psi_i$, which is proportional to
the pressure field within the layer, and such that the horizontal
velocities ${\bf v_i} = (u_i,v_i)$ within the layer verify
$u_i=-\frac{\partial
\psi_i}{\partial y}$ and $v_i=\frac{\partial \psi_i}{\partial x}$. In
our equations, the coordinate directions $x$ and $y$ will be
oriented along the northward and the eastward directions,
respectively.

More specifically we use a {\it two-layer quasigeostrophic model
on a beta plane and over topography}. An {\it eddy-viscosity} and
a {\sl bottom friction} terms are also included. The equations
defining the dynamics of the stream function of both layers are
\begin{eqnarray}
\frac{D_1}{Dt} \left[ \nabla^2 \psi_{1} + f+
\frac{\psi_{2}-\psi_{1}}{R_1^2} \right]& = &\nu \nabla^4 \psi_{1},
\label{capa1}
\\
 \frac{D_2}{Dt}\left[  \nabla^2\psi_{2}+f
+\frac{\psi_{1}-\psi_{2}}{R_2^2}+ f_0\frac{\tau}{H_2} \right] &=&
\nu
\nabla^4 \psi_{2}-C_b\nabla^2 \psi_{2}, \label{capa2}
\end{eqnarray}
\noindent where the subscript $i=1$ ($2$) refers to the upper (bottom)
layer, $\frac{D_i}{Dt}= \frac{\partial}{\partial
t}+J(\psi_i,\cdot)$, $\psi_i(x,y,t)$ is the layer stream function,
$\tau (x,y)$ is the bottom topography, $R_1=\frac{N_1 H_1}{f_0}$,
$R_2=\frac{N_2 H_2}{f_0}$, $N_1^2=\frac{g \delta \rho }{H_1\rho_2
}$, $N_2^2=\frac{g \delta
\rho}{H_2\rho_2}$, where $\delta \rho =\rho_2 - \rho_1$. $g$ the
gravitational acceleration, $H_i$ is the mean thickness
of the layer
$i$,
 and $f$ is the {\it Coriolis
parameter}, which in the beta-plane approximation depends on the
latitude as $f=f_0 +
\beta y$. $\nu$ is the
eddy-viscosity and $C_b$ is the coefficient describing friction
with the bottom of the sea. The Jacobian operator $J(A,B)$ is
defined as:
\begin{equation}
J(A,B)=\frac{\partial A}{\partial x}\frac{\partial B}{\partial y}-
\frac{\partial A}{\partial y} \frac{\partial B}{\partial x}.
\label{jacobiano}
\end{equation}

More details about quasigeostrophic dynamics can found for example
in Refs. ~\cite{Ped} and ~\cite{Le}. A more explicit way to write
our equations (\ref{capa1}) and (\ref{capa2}) is
\begin{eqnarray}
\frac{\partial \nabla^2 \psi_1}{\partial t}
+ \frac{1}{R_1^2}\frac{\partial (\psi_2-\psi_1)}{\partial t}
+
&&\frac{\partial\nabla^2 \psi_1}{\partial x} \frac{\partial \psi_1}
{\partial y}-
\frac{\partial\nabla^2 \psi_1}{\partial y} \frac{\partial \psi_1}{\partial x}
  \nonumber \\
- \frac{1}{R_1^2} \left(  \frac{\partial \psi_1} {\partial
x}\frac{\partial \psi_2}{\partial y}+
\frac{\partial \psi_1}{\partial y}\frac{\partial \psi_2}{\partial x}  \right)
+   && \beta \frac{\partial \psi_1}{\partial x}= \nu \nabla^4  \psi_1,
\label{capaup} \\
\frac{\partial \nabla^2 \psi_2}{\partial t}
+ \frac{1}{R_2^2}
\frac{\partial (\psi_1-\psi_2)}{\partial t}
+ \nonumber  && \frac{\partial\nabla^2 \psi_2}
{\partial x} \frac{\partial \psi_2}{\partial y}-
 \frac{\partial \psi_2}{\partial x}\frac{\partial \nabla^2 \psi_2}
 {\partial y}
 \\ \nonumber
-\frac{1}{R_2^2} \left(  \frac{\partial \psi_2} {\partial
x}\frac{\partial \psi_1}{\partial y}+
\frac{\partial \psi_2}{\partial y}\frac{\partial \psi_1}{\partial x}  \right)
+  && \frac{f_0^2}{gH_2}    \left(  \frac{\partial \psi_2}
{\partial x}\frac{\partial \tau}{\partial y} - \frac{\partial
\psi_2}{\partial y}\frac{\partial \tau}{\partial x}  \right) + \beta
\frac{\partial \psi_2}{\partial x}=
\nonumber
\\
\nu \nabla^4
\psi_2 - C_b\nabla^2 \psi_{2}
\label{capadown}
\end{eqnarray}

In this model the stream function of the upper layer is
$\psi_1(x,y,t)=\frac{g}{f_0} h(x,y,t) $ where $h(x,y,t)$ is the
height of the sea surface over the point $(x,y)$ at time $t$. This
last quantity is the one linked to the satellite observations on
which we have performed the POD. It is important to note that in
equations (\ref{capaup}) and (\ref{capadown}) we have not
considered an annual forcing term which accounts for the the
seasonal variability of the Algerian Current. This is a very
important fact for the next step in our approach, the projection
onto a particular Galerkin basis determined from the observations.
We assume the following ansatz:
\begin{eqnarray}
\psi_1(x,y,t)&=&<\psi(x,y,t)>+\psi'(x,y,t)
\nonumber \\
    &=&\frac{g}{f_0}(<h(x,y,t)> +\hat a_3(t)\phi_3(x,y)+
	\hat a_4(t) \phi_4(x,y)),
    \label{ansatz1}
\end{eqnarray}
where the temporal coefficients $\hat a_3(t)$ and $\hat a_4(t)$
will be  calculated in the following sections and, 
finally,  compared with the POD temporal coefficients
$a_3(t)$ and $a_4(t)$.
 With this we assume that the
stream function $\psi_1$ (which is proportional to the height) of
the upper layer is decomposed in its temporal mean
$<\psi(x,y,t)>=\frac{g}{f_0}<h(x,y,t)>$, 
accounting for the annual mean flow, and a
perturbation $\psi'(x,y,t) =
\frac{g}{f_0} (\hat a_3(t)\phi_3(x,y)+\hat a_4(t)\phi_4(x,y))$, which models
the mesoscale processes \cite{BK}. The perturbation basis for the
height $\{\phi_3(x,y),\phi_4(x,y)\}$ is what we have obtained from
the POD analysis of the data, and the way to calculate the annual
mean will be detailed at the end of this Section.

As we have no real measure for the bottom layer (the satellite
sensors get data just from the sea surface), we need to make some
additional hypothesis in our model. We propose the following
ansatz for the expansion of the bottom layer's stream function
\begin{eqnarray}
\psi_2(x,y,t)&=& -Uy+Vx+ \frac{g}{f_0}(b_3 (t)\phi_3(x,y)+b_4(t)
\phi_4(x,y)), \label{ansatz2}
\end{eqnarray}
\noindent where $U$ and $V$ are parameters of our model and
simulate the eastward and northward velocity, respectively, of the
bottom layer flow. The physical  meaning of (\ref{ansatz2}) is
that the perturbations from the mean flow for the bottom layer are
generated by the same basis functions as the upper one though
with, obviously, different temporal coefficients.
%Also, it is implicitly assumed in \ref{ansatz2} that the
%influence on the upper layer of the perturbations of the mean flow
%in the bottom layer is negligible.
The most important feature in the former ansatz is the mean flow,
parameterized with $U$ and $V$. The values of these parameters not
only determine the intensity of the mean flow but also, and most
importantly, its direction and sense. Discussions about the
r\^{o}le of the different values of $U$ and $V$ will be given in
the next section.

Projecting expansions (\ref{ansatz1}) and (\ref{ansatz2}) over
equations (\ref{capaup}) and (\ref{capadown}) and using the
orthogonality relations of the POD basis $\phi_i(x,y)$, we obtain the
evolution equations for the coherent structure's temporal
amplitudes $\hat a_i$ and $b_i$ ($i=3,4$),
\begin{eqnarray}
\frac{d\hat a_i}{dt}&=&p_{1,i}\hat a_3^2+p_{2,i}\hat a_4^2+p_{3,i}\hat a_3
+p_{4,i}\hat a_4+p_{5,i}\hat a_3
\hat a_4 \nonumber
\\&+& p_{6,i}b_3^2+p_{7,i}b_4^2  +p_{8,i}b_3
+p_{9,i}b_4+p_{10,i}b_3b_4 +p_{11,i}, \nonumber \\
\frac{db_i}{dt}&=&q_{1,i}\hat a_3^2+q_{2,i}\hat a_4^2
+q_{3,i}\hat a_3+q_{4,i}d_2+q_{5,i}\hat a_3
\hat a_4 \nonumber \\ &+&
q_{6,i}b_3^2+q_{7,i}b_4^2+q_{8,i}b_3+q_{9,i}b_4+q_{10,i}b_3b_4
+q_{11,i}.
\label{aee}
\end{eqnarray}

 The
coefficients $p_{k,i} $ and $q_{k,i}$ ($k=1,..,11 $ and $ i=3,4$)
are real numbers that depend on the parameters of the model and on
integrals containing $< \psi
>$, $\phi_3(x,y)$, $\phi_4(x,y)$, the and their derivatives.
Their explicit expressions are quite complex and have been
obtained by computer algebraic manipulation. We do not write down
here all these involved expressions. Just to give an example of
them, in Appendix A we display the mathematical expression for
$q_{10, 2}$.

We finally proceed to explain how to obtain the mean flow
$<\psi(x,y,t)>$, without which the coefficients in (\ref{aee})
remain undetermined. Some manipulations of the data are needed
because of the bad spatial resolution of the available mean field.
We recall that we are dealing with Sea Level Anomaly (SLA) data
obtained from the two altimetric missions ERS-1 and T/P. 
 These are conveniently
treated to obtain regular maps in space and time every 10 days and
on a $0.2^{o}$ regular grid. These SLA are relative to the annual
mean sea level and, therefore, we need this annual mean to obtain
the total height of the sea surface and thus be able to make
relation between the empirical eigenfunctions $\phi_3$, $\phi_4$
and the dynamic variable $\psi_1$ of the quasigeostrophic model
(see Eq.~\ref{ansatz1}). Unfortunately, the annual mean sea level
data are not manipulated to improve their resolution, and we have
just the T/P data, of a very coarse resolution (around $2.8 ^{o}$)
to calculate this annual mean. Therefore, we need to interpolate
these to a $0.2^{o}$ regular grid and then to add the resulting
annual mean to the SLA data, in order to obtain a consistent SSH
field.
But, all these manipulations are, at the end, manifesting when
we  solve eq.(\ref{aee})
  in such a way that $\hat a_3(t)$ and $\hat a_4(t)$ have a 
 nonzero average, in contrast with the POD  temporal modes,
  $a_3(t)$ and $a_4(t)$, calculated from data,
 i.e.,
\begin{equation}
 \frac{1}{T}\int_0^T \hat a_i (t) dt \ne 0, \ \ i=3,4.
 \end{equation}

  In order to heal this, we
 proceed with an assimilation-like approach:
thus we modify the annual mean sea level, which has been
obtained interpolating
the T/P data, by adding this nonzero average, i.e.,
\begin{eqnarray}
h_m(x,y)=h_m^{T/P}(x,y)+\frac{f_0}{g}\frac{1}{T} \int_0^T \hat a_3
(t) (\phi_3(x,y,t)+ \hat a_4 (t) \phi_4(x,y,t)) dt,
\end{eqnarray}
being $h_m(x,y)$ the new  annual mean height
 and $h_m^{T/P}(x,y)$
is the interpolated T/P annual mean height. Finally, and with high
numerical accuracy, the new temporal modes obtained from 
eq. (\ref{aee})
have now temporal zero
average.

Summing up, the data we are using along this paper are obtained by
adding to the SLA data the above calculated $h_m(x,y)$ field. It is
important to note that because in the POD analysis we substract
the mean field of the data, the qualitative  features of the
analysis  are similar if we analyse the SLA data or the SLA plus the
$h_m$ field. Nonetheless, the quantitative differences are
not negligible at all at the level of the dynamical system
(\ref{aee}).

The four-dimensional dynamical system (\ref{aee}), now fully
defined, is the desired low-dimensional approximation aimed to
describe the coherent structures in our data set. In the next
Section we show that the dynamics of the observed coherent eddy is
recovered from (\ref{aee}).

\section{Numerically generated coherent structure dynamics}
\label{numerical}

We now proceed to integrate the equations (\ref{aee}). First,
  typical values for the parameters of the quasigeostrophic
model (\ref{capaup}) and (\ref{capadown}) are needed. At
mid-latitudes, adequate values for the parameters giving the
Coriolis force are $\beta = 1.0\ 10^{-11}
\  m/s $ and $f_0=10^{-4}s^{-1}$ \cite{Ped,Benoit}. In addition,
for the Algerian Current area the values $\rho_{1} =1025
\ kg/m^3$ and $\rho_2=1029 \ kg/m^3 $ are adequate for the densities of the
upper and bottom layer and $H_1=300 \ m $, $H_2=3500 \ m $ for
their mean heights. 
The bottom topography
$\tau (x,y)$
are real data obtained from the data basis at the URL
'http://modb.oce.ulg.ac.be/Bathymetry.html'.
For the friction parameter with the bottom
topography we take a value of $C_b=1.5 \ 10^{-6}$. A typical value
for the eddy viscosity at the scales we are working is $\nu =200 \
m^2/s$.
 The remaining parameters are the geostrophic
velocities of the bottom layer. This is a difficult task since
nobody really knows what happens in the deep waters of the
Algerian Current. Anyway, recent results obtained by the PRIMO-1
experiment in the channel of Sardinia \cite{Bou1} seem to indicate
that typical velocities for the deep layer are rather weak (of the
order of 1 $cm/s$), though it is not clear if the {\it deep layer
flow} (dlf) is westward or eastward. Therefore we assume dlf
following the upper layer, this is northeastward, with typical
values of: $U=5 \ cm/s$ and $V=5 \ cm/s$. In the next Section,
other values of $U$ and $V$ will be discussed.

Fig.~\ref{fig:temporalmodesNUM} shows $\hat a_3(t)$
and $\hat a_4(t)$
obtained by integrating (\ref{aee}) with a fourth order
Runge-Kutta method. The periodicity of both is evident, being the
period of around 5 moths, in good agreement with the observed one.
If some annual forcing would be added to our model, this period
would probably lock to the semiannual harmonic, still improving
the agreement. The shape of the oscillation in the calculated
evolution is much more regular than the experimental one, as
expected from a clean simulation versus noisy data. In
Fig.~\ref{fig:movie} we show a temporal sequence of the coherent
structure dynamics given by $\hat a_3(t)\phi_3(x,y)+\hat a_4(t)\phi_4(x,y)$
from the calculated temporal modes. An eddy appears next to the
coast and moves northeastward, the process being repeated five
months later. This is fully consistent with the observed data and
confirms the success in our task of obtaining a low-dimensional
reduced model. Sensibility of the results to variations in the
values chosen for the parameters is discussed in terms of a
bifurcation analysis in the next Section.

\section{Bifurcation analysis}
\label{ba}

The value of the eddy viscosity $\nu$ is somehow arbitrary since
it would depend on the scale of observation. Thus, a discussion of
the variations in model behavior as $\nu$ varies is in order. We
analyze numerically our system of four ODE's (\ref{aee}) with the
help of the software package {\it Dstool } \cite{dstool} which
integrates the system with a fourth order Runge-Kutta. We change
the eddy-viscosity parameter $\nu$, which is proportional to the
inverse of the Reynolds number, and observe the bifurcation
behavior. The rest of the parameters take the same values as in
the former section. It should be noted that in principle the
empirical eigenfunctions would vary in a system with varying
$\nu$, but since we just have experimental data for the actual
value of the eddy viscosity in the Ocean at the observed scales,
we keep the parameters in (\ref{aee}) as determined from the POD
of the observed data.  The bifurcation diagram is outlined as
follows:
\begin{itemize}
\item For $\nu \ge 212 \ m^2/s$  there are six
fixed points. Two of them are stable and the rest are unstable.

\item When $\nu =212 \ m^2/s$ a 
Hopf bifurcation occurs. One of the stable
fixed points (the one localized at the origin) gets unstable by
decreasing $\nu$ and a limit cycle appears surrounding it. The
limit cycle persists for all the values of the viscosity smaller
than $212 \ m^2/s$. 
The system undergoes no new bifurcations by decreasing
the viscosity parameter.

\end{itemize}

The main dynamical feature in our model is thus the existence of a
Hopf bifurcation which gives birth to a limit cycle for a long
range of $\nu $ values, including the physical ones at the scales
we are working. All these limit cycle solutions give rise, after
reconstruction of the coherent structure with the help of the
empirical eigenfunctions to traveling wave patters with a period
of around six months. In particular, the moving eddy identified in
section \ref{numerical} is just one of these solutions. Moreover,
the rest of the  fixed points in the second regime, i.e. when $\nu
\le 212 \ m^2/s$, seem 
to have no physical significance as their basins of
attraction correspond to very high values of the initial condition
for $\hat a_i (t)$
($i=3,4$), i.e., high values of the sea surface
height. When the eddy-viscosity is too large, the system evolves
towards a stable fixed point, with no moving coherent structures,
as expected on physical grounds.

To give a stronger support to the former analysis, we have tested
our ODE system with other values of the dlf velocity.
 We have
observed numerically the following behaviour of the system:
\begin{itemize}

\item High values of $U$ or $V$, $\sim 10 \ cm/s$,
produce diverging solutions, 
 with no fixed points for any
value of $\nu$. Very low values ($\sim 10^{-2}
\ cm/s$) give rise also to unbounded solutions for any initial
condition in some range of high viscosity values (low Reynolds
number), which is not reasonable on physical grounds.

\item Changing sign in $U$, that is, assuming
a westward flow in the bottom layer, does not change considerably
the bifurcation diagram, but the amplitude of the limit cycle is
too big when reasonable values of the viscosity (around $200 \
m^2/s$) are used. On the contrary, changing sign in $V$ or in $U$
and $V$ simultaneously gives rise to a
 ODE system where all the  solutions are unbounded.

\end{itemize}

We think these are enough reasons supporting the chosen direction
and magnitude of the velocity of the dlf, that is, a northeastward
direction and typical values for the horizontal velocites around
$5 \ cm/s$.

\section{Summary}

In this article we have used the Proper Orthogonal Decomposition
to obtain a low-dimensional dynamical description of coherent
structures observed in satellite data of a region of the
Mediterranean Sea. First, analysis of the altimetric satellite
data via the POD allows the identification of a moving vortex in
the ocean surface. Second, projection of a two-layer
quasigeostrophic model onto the empirical basis, together with
some physical assumptions on the unobserved part of the sea, allow
the construction of a fourth-order dynamical system that gives a
reasonable description of the dynamics of the coherent structure,
in particular its period and amplitude. It is remarkable that a
crude PDE model, and noisy data, can be merged to obtain an
efficient reduced model. We expect our general methodology would
be of use in other complex environmental fluid dynamics problems.

\section{Appendix}
\label{appendix}

If we denote the scalar product by $\{ \phi_3 , \phi_4
\}=\int_A
 \phi_3(x,y) \phi_4(x,y) dx dy$, and we define also:
 \begin{eqnarray}
c_{k,l}&=&-\{\triangle \phi_k,\phi_l\},  \\ e_{k,l}&=&\{-\beta
\frac{\partial \phi_k}{\partial x} +\nu \nabla^{4} \phi_k
-\frac{g}{f_0}\frac{\partial<\psi>}{\partial x}\frac{\partial
\triangle \phi_k}{\partial y}+\frac{\partial \triangle
<\psi>}{\partial y}\frac{\partial \phi_k}{\partial x}\nonumber \\
& -& \frac{\partial <\psi>}{\partial y}\frac{\partial \triangle
\phi_k}{\partial x} - \frac{\partial \triangle <\psi>}{\partial
x}\frac{\partial \phi_k}{\partial y}-\frac{g}{f_0 R_1^{2}}(U
\frac{\partial \phi_k}{\partial x}-V \frac{\partial
\phi_k}{\partial y} ),\phi_l\},\\
 p_{k,l}&=&\frac{g}{f_0 R_1^{2}}\{\frac{\partial <\psi> }{\partial
y} \frac{\partial \phi_k}{\partial x} -\frac{\partial
<\psi>}{\partial x} \frac{\partial \phi_k}{\partial y} , \phi_l\},
\\ f_{k,l}&=&\{-\frac{f_0}{H_2}(\frac{\partial \tau}{\partial y}
\frac{\partial \phi_k}{\partial x} - \frac{\partial \tau}{\partial
x} \frac{\partial \phi_k}{\partial y} ) - \beta \frac{\partial
\phi_k}{\partial x} - C_b \triangle \phi_k +\nu \nabla^{4}\phi_k
\nonumber \\& - & \frac{g}{f_0}(\frac{\partial \triangle
\phi_k}{\partial y} V \nonumber  - \frac{\partial \triangle
\phi_k}{\partial x} U) - \frac{g}{f_0 R_2^{2}}(\frac{\partial
<\psi>}{\partial y } \frac{\partial \phi_k}{\partial x} -
\frac{\partial <\psi>}{\partial x} \frac{\partial \phi_k}{\partial
y}),\phi_l\},\\
 h_{k,l}&=&
-\frac{g}{f_0 R_2^{2}}\{V \frac{\partial \phi_k}{\partial y} - U
\frac{\partial \phi_k}{\partial x}, \phi_l\},\\
 q_{1,l}&=&
\{-\frac{g}{f_0}\frac{\partial <\psi>}{\partial x}\frac{\partial
\triangle <\psi>}{\partial y}+\frac{g}{f_0}\frac{\partial
<\psi>}{\partial y}\frac{\partial \triangle <\psi>}{\partial
x}-\frac{g}{f_0 R_1^{2}}U\frac{\partial <\psi>}{\partial x}
\nonumber \\ & + & \frac{g}{f_0 R_1^{2}}V\frac{\partial
<\psi>}{\partial y}-\beta \frac{\partial<\psi>}{\partial x}+\nu
\nabla^{4} <\psi>, \phi_l\},
\\ q_{2,l}&=&\{-\frac{g}{f_0
R_2^{2}}V\frac{\partial<\psi>}{\partial y}+\frac{g}{f_0
R_2^{2}}U\frac{\partial<\psi>}{\partial x}+\frac{f_0}{H_2}U
\frac{\partial \tau}{\partial x} \nonumber \\
& - & \frac{f_0}{H_2}V \frac{\partial
\tau}{\partial y} -\beta V , \phi_l\},\\ d_{1,l}&=&
-\frac{g}{f_0}\{\frac{\partial \phi_3}{\partial x} \frac{\partial
\triangle \phi_3 }{\partial y}-\frac{\partial \phi_3}{\partial y}
\frac{\partial \triangle \phi_3 }{\partial x},\phi_l\},\\
d_{2,l}&=& -\frac{g}{f_0}\{\frac{\partial \phi_4}{\partial x}
\frac{\partial \triangle \phi_4 }{\partial y}-\frac{\partial
\phi_4}{\partial y} \frac{\partial \triangle \phi_4 }{\partial
x},\phi_l\},\\ d_{3,l}&=&\{-\frac{g}{f_0}(\frac{\partial
\phi_3}{\partial x} \frac{\partial \triangle \phi_4 }{\partial y}
- \frac{\partial \phi_3}{\partial y} \frac{\partial \triangle
\phi_4 }{\partial x}+\frac{\partial \phi_4}{\partial x}
\frac{\partial \triangle \phi_3 }{\partial y} - \frac{\partial
\phi_4}{\partial y} \frac{\partial \triangle \phi_3 }{\partial
x},\phi_l \},
 \end{eqnarray}
with $j,l=3,4$, we  finally obtain,
\begin{eqnarray}
q_{10,2}&=&\frac{1}{k_1} \left(
{{{R_{{1}}}^{2}{R_{{2}}}^{4}d_{{3,3}}c_{{4,4}}c_{{3,4}}} }- {
{{c_{{3,3}}}^{2}d_{{3,4}}{R_{{2}}}^{4}{R_{{1}}}^{2}} }  +
 {{R_{{2}}}^{4}d_{{3,3}}c_{{3,4}}}+ {c_{{1,2
}}{R_{{1}}}^{4}d_{{3,4}}{R_{{2}}}^{2}c_{{4,3}}}
\right.
\nonumber
\\
&+& {{R_{{1}}}^{2}{R_{{2}}}^{2}c_{{3,4}}d_{{3,3}}}-
{c_{{3,3}}{R_{{1}}}^{2}d_{{3,4}}{R_{{2}}}^{2}}  - {c_{{4,4}}{R_{{1
}}}^{4}c_{{3,3}}d_{{3,4}}{R_{{2}}}^{2}}-
{c_{{4,4}}{R_{{1}}}^{2}c_{{3,3}}{R_{ {2}}}^{4}d_{{3,4}}}
\nonumber
\\
&-& {c_{{3,3}}{R_{{2}}}^{4}d_{{3,4}}}+
{c_{{3,4}}{R_{{1}}}^{4}d_{{3,4}}{R_{{2}}}^{4}c_{{4,3}}c_{{3,3}}}
 -  {{c_{{3,4}}}^{2}{R_{{1}}}^{4}c_{{4,3}}{
R_{{2}}}^{4}d_{{3,3}}}+
{c_{{4,4}}{R_{{1}}}^{4}c_{{3,3}}{R_{{2}}}^{4}c_{{3,4} }d_{{3,3}}}
\nonumber
\\
&-&
\left.
 {c_{{4,4}}{R_{{1}}}^{4}{c_{{3,3}}}^{2}d_{{3,4}}{R_{{2}}}^{4}}
+   {c_{{3,3 }}{R_{{2}}}^{4}c_{{3,4}}{R_{{1}}}^{2}d_{{3,3}}}
           \right),
\end{eqnarray}

where
\begin{eqnarray}
k_1 &=& -2 c_{
{4,3}}{R_{{2}}}^{2}c_{{3,4}}{R_{{1}}}^{2}+{c_{{4,4}}}^{2}{R_{{2}}}^{2}
{R_{{1}}}^{4}c_{{3,3}}+{c_{{4,4}}}^{2}{R_{{2}}}^{4}{R_{{1}}}^{2}c_{{3,
3}}+{c_{{4,4}}}^{2}{R_{{2}}}^{4}{R_{{1}}}^{4}{c_{{3,3}}}^{2}
\nonumber
\\
&+& c_{{4,4}}
{R_{{1}}}^{4}c_{{3,3}}-c_{{4,4}}{R_{{1}}}^{4}c_{{4,3}}{R_{{2}}}^{2}c_{
{3,4}}+2
c_{{4,4}}{R_{{1}}}^{2}c_{{3,3}}{R_{{2}}}^{2}+c_{{4,4}}{R_{{1
}}}^{4}{c_{{3,3}}}^{2}{R_{{2}}}^{2}-c_{{4,3}}{R_{{2}}}^{2}c_{{3,4}}{R_
{{1}}}^{4}c_{{3,3}}
\nonumber
\\
&+&{c_{{4,3}}}^{2}{R_{{2}}}^{4}{c_{{3,4}}}^{2}{R_{{1}
}}^{4}-c_{{4,3}}{R_{{2}}}^{4}c_{{3,4}}{R_{{1}}}^{2}c_{{3,3}}-c_{{4,4}}
{R_{{2}}}^{4}c_{{4,3}}c_{{3,4}}{R_{{1}}}^{2} +
c_{{4,4}}{R_{{2}}}^{4}c_{
{3,3}}+c_{{4,4}}{R_{{2}}}^{4}{c_{{3,3}}}^{2}{R_{{1}}}^{2}
\nonumber
\\
&-& c_{{3,4}}{R_ {{1}}}^{4}c_{{4,3}}-2
c_{{4,4}}{R_{{2}}}^{4}c_{{3,4}}{R_{{1}}}^{4}c_{
{4,3}}c_{{3,3}}-c_{{3,4}}{R_{{2}}}^{4}c_{{4,3}}
\end{eqnarray}

\section*{Acknowledgments}

We thank useful discussions with J. M. Pinot and J. Tintor\'e.
Financial support from CICyT
projects MAR95-1861 and MAR98-0840 is also acknowledged.
The data used in this work are ESA's ERS-1 and ERS-2 data,
CNES/NASA TOPEX/POSEIDON data and MSLA (Maps of Sea Level 
Anomalies) products \cite{Ay}.

\newpage

\begin{figure}
\epsfig{file=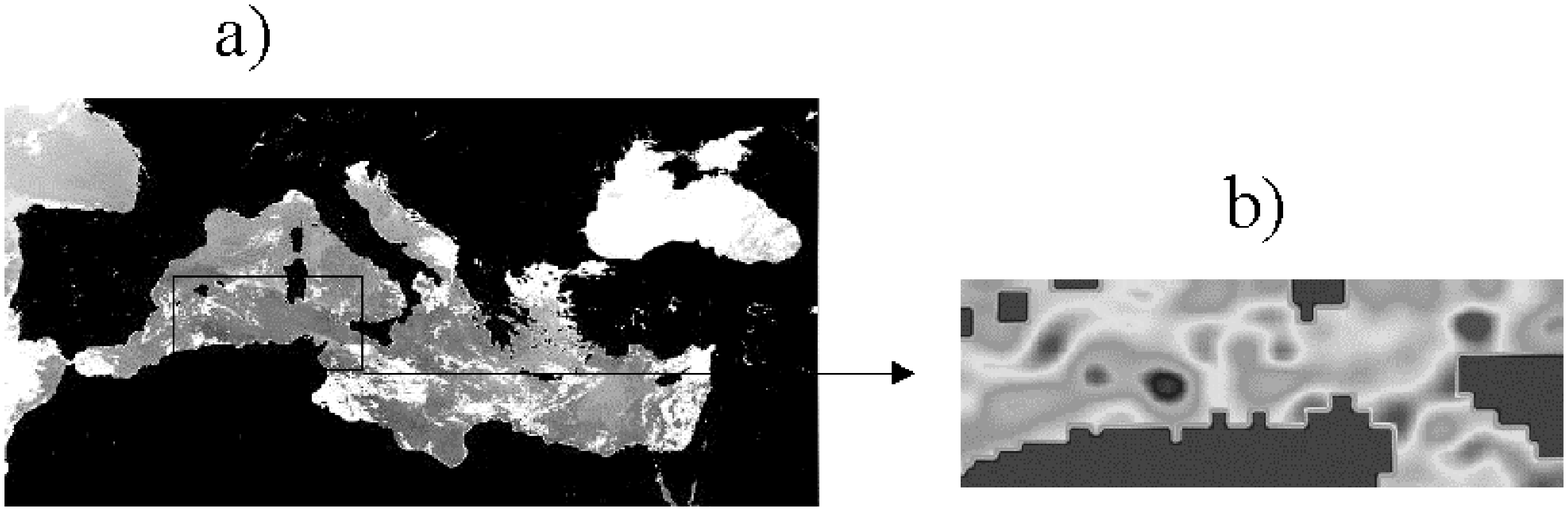,width=.8\linewidth}
\caption{a) 
A map of the Mediterranean Sea.
In the small box we show the area
under study. b) Shows one of the altimetry images obtained from the 
satellite.
}
\end{figure}

\begin{figure}
\epsfig{file=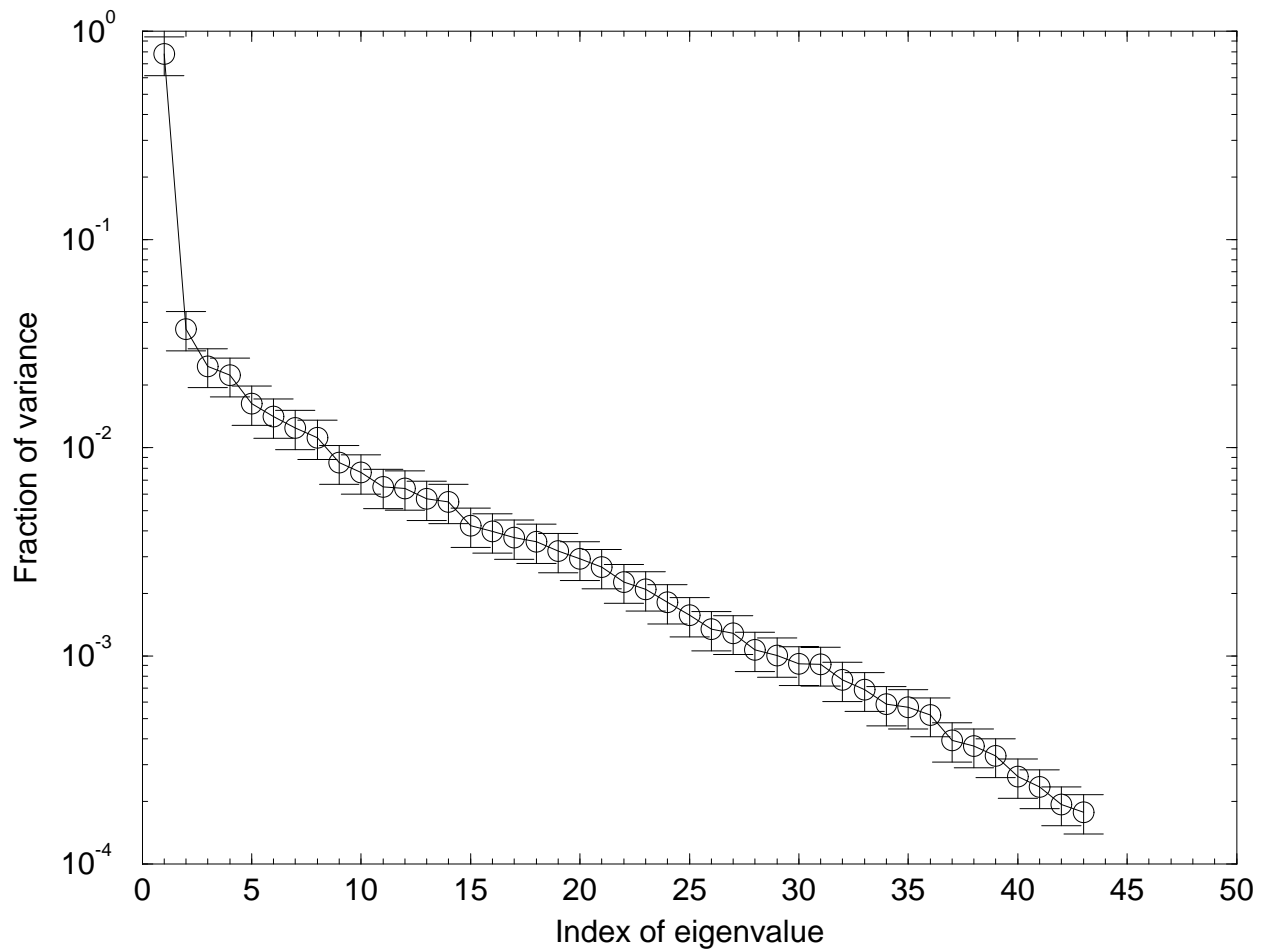,width=.8\linewidth,angle=-90}
\caption{Linear-log plot of the fraction of variance of
the eigenvalues in terms of their index.}\label{fig:fullspectrum}
\end{figure}

\begin{figure}
\epsfig{file=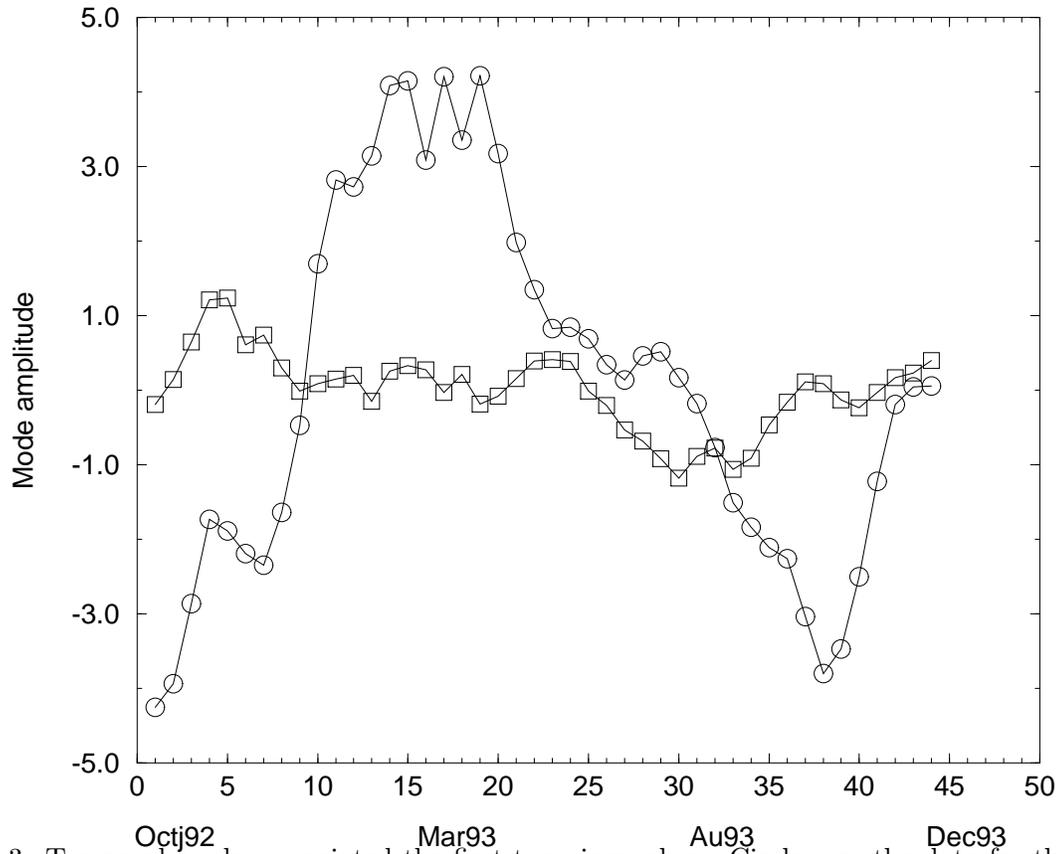,width=.8\linewidth,angle=-90}
\caption{Temporal modes associated the first two eigenvalues.
Circles are the data for the first and squares for the second.}
\label{fig:anualmodes}
\end{figure}

\begin{figure}
\epsfig{file=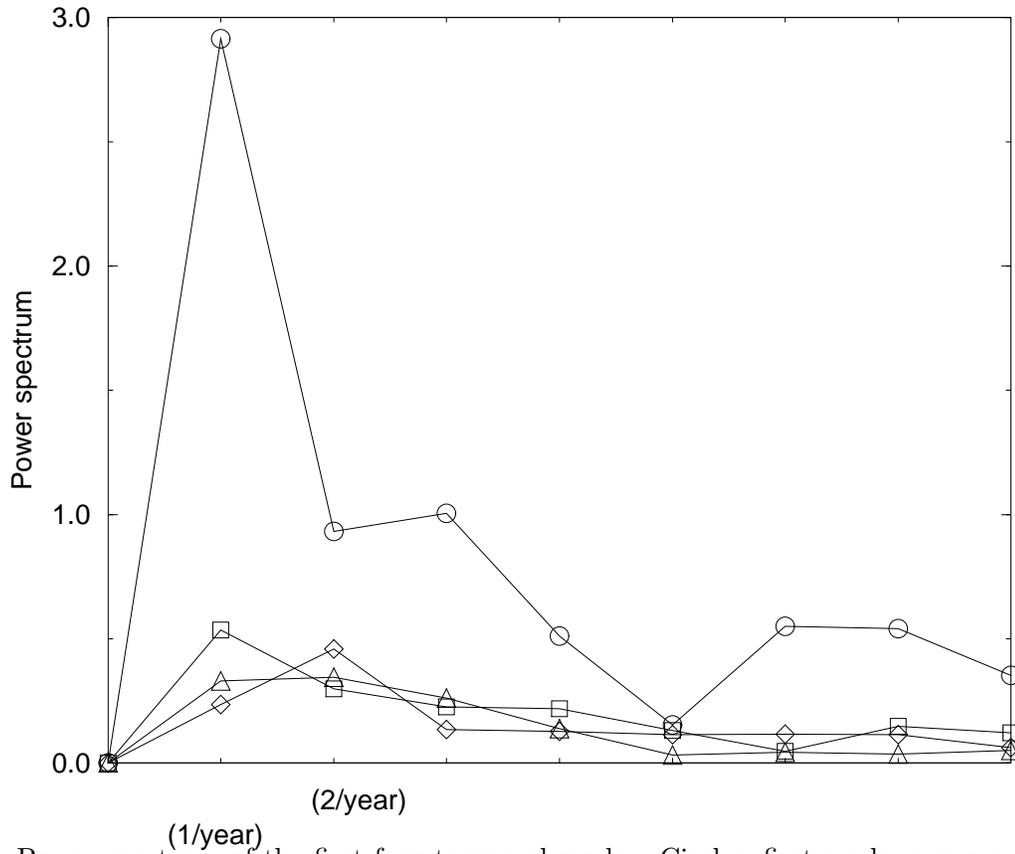,width=.8\linewidth,angle=-90}
\caption{Power spectrum of the first four temporal modes. Circles,
first mode; squares, second mode; triangles, third mode; and
diamonds, fourth mode. }
\label{fig:power}
\end{figure}

\begin{figure}
\epsfig{file=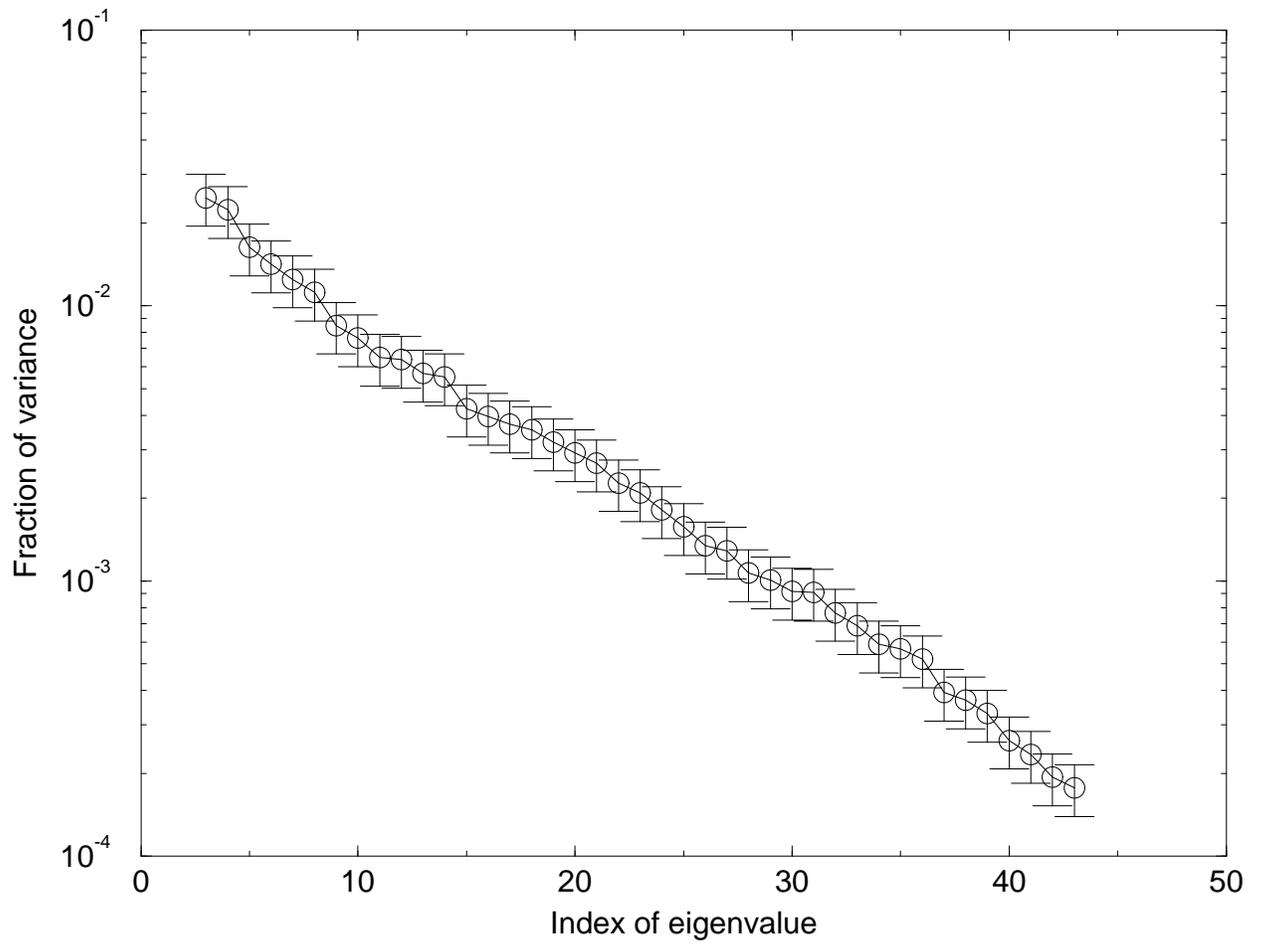,width=.8\linewidth,angle=90}
\caption{Linear-log plot showing the fraction of variance contained in
the different eigenvalues starting from the third one.}
\label{fig:partialspectrum}
\end{figure}

\begin{figure}
\epsfig{file=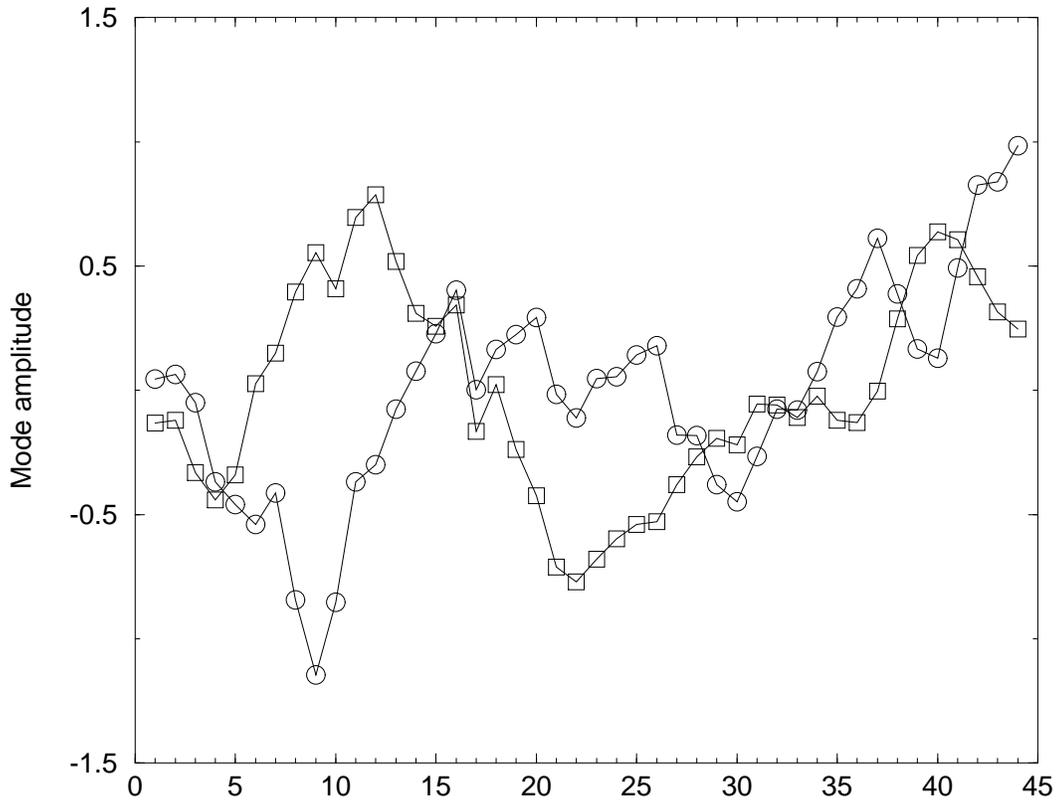,width=.8\linewidth,angle=-90}
\caption{Temporal modes for the third and fourth eigenfunctions.
Squares, third mode;
circles, fourth mode.
}
\label{fig:temporalmodes34}
\end{figure}

\begin{figure}
\epsfig{file=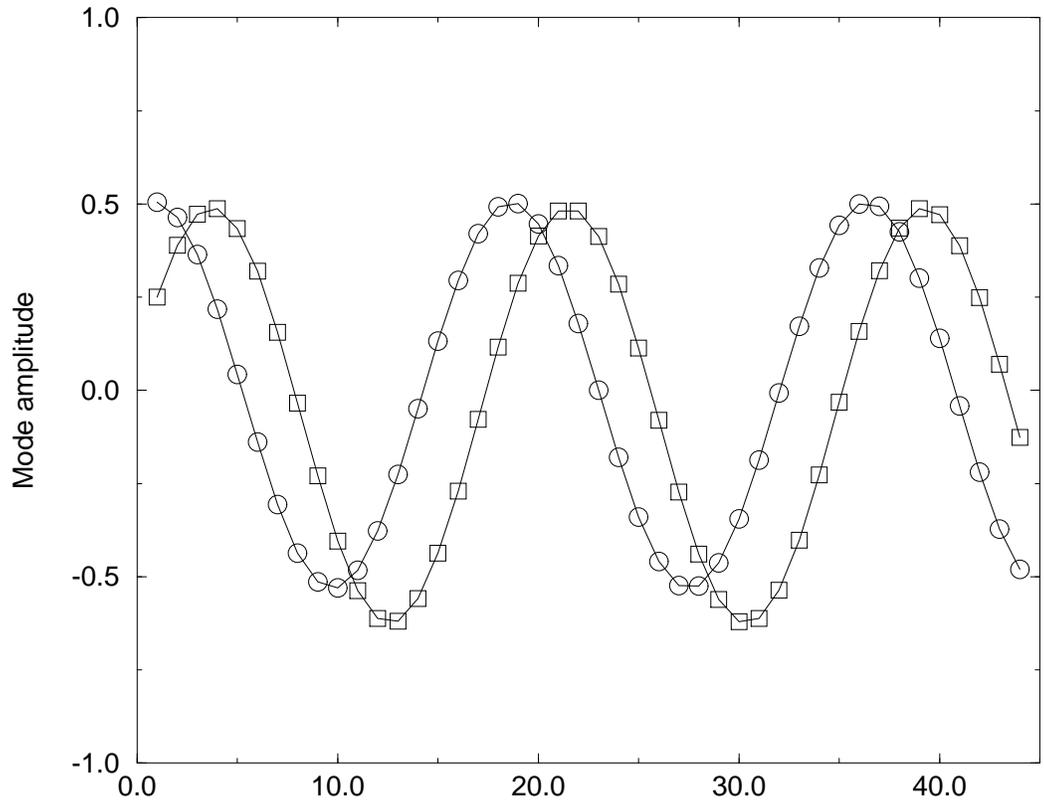,width=.8\linewidth,angle=-90}
\caption{Temporal evolution of the
third and fourth temporal modes obtained by numerical integration
Eq.~\ref{aee}}
\label{fig:temporalmodesNUM}
\end{figure}

\begin{figure}
\epsfig{file=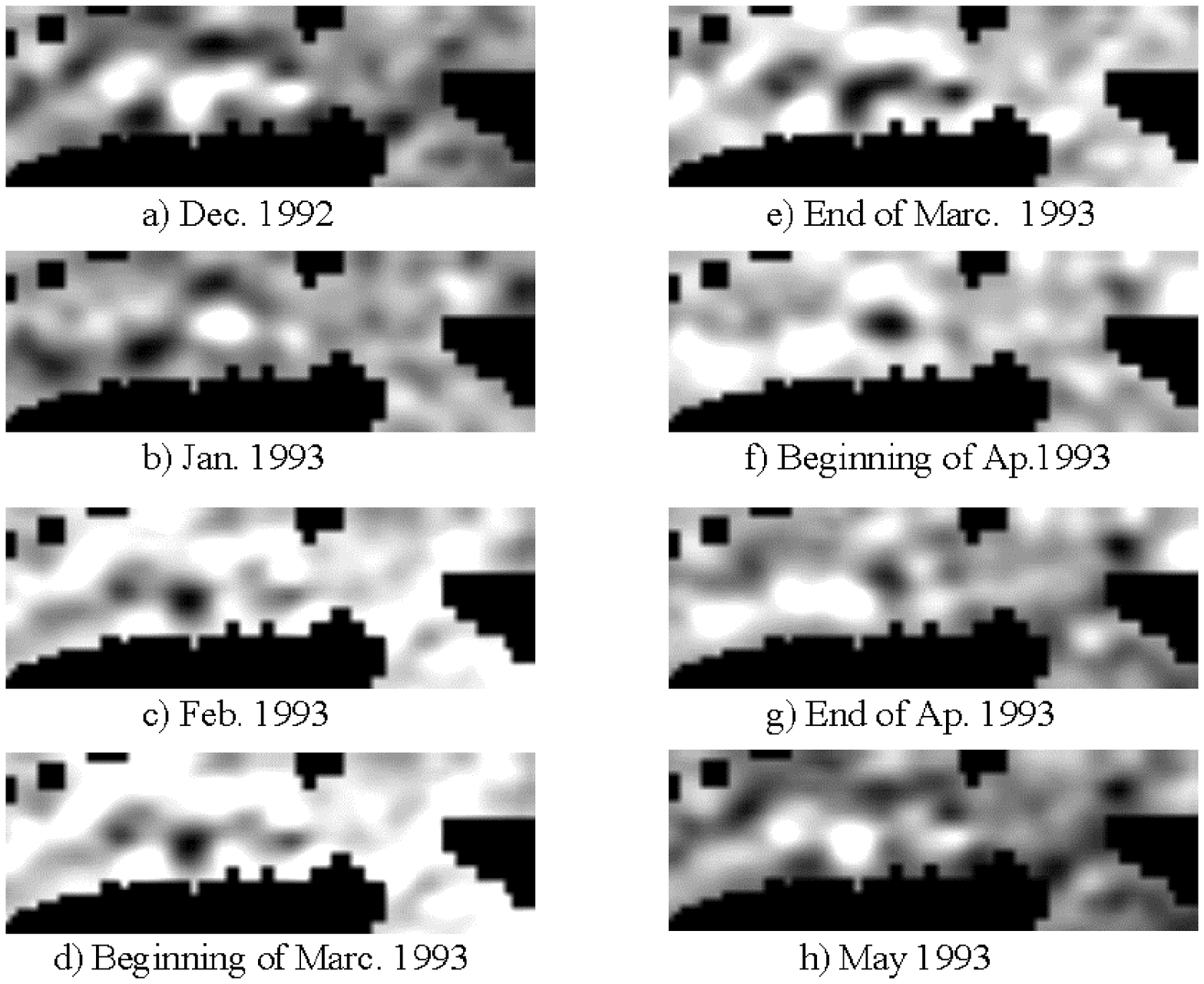,width=1.2\linewidth}
\caption{Numerical reconstruction of vortex shedding and motion
near the Algerian coast.
After h) the same sequence is repeated.
The  lighter
area (higher altimetric values) next to the coast
corresponds
to an anticyclonic  vortex (the moving coherent
structure under study).
In the figure, it always appears with its corresponding
cyclonic vortex (darker area).
}
\label{fig:movie}
\end{figure}


\begin{thebibliography}{00}

% \bibitem{label}
% Text of bibliographic item

\bibitem{HLB}
P. Holmes, J.L. Lumley and G. Berkooz, {\it Turbulence, Coherent
Structures, Dynamical Systems and Symmetry}, Cambridge University
Press, Cambridge, 1996; P. Holmes, J.L. Lumley, G. Berkooz, J.C.
Mattingly, and R.W. Wittenberg, Phys. Rep. {\bf 287} (1997), 337;
G. Berkooz, P. Holmes, and J.L. Lumley, Annu. Rev. Fluid Mech.
{\bf 25} (1993), 539.

\bibitem{lu}
J.L. Lumley, {\it Stochastic Tools in Turbulence }. Academic
Press, New York, 1971.

\bibitem{Preisen}
R. W. Preisendorfer, {\it Principal Component Analysis in
Meteorology and Oceanography}. Elsevier, Amsterdam, 1988.

\bibitem{Palacios96}
A. Palacios, D. Armbruster, E.J. Kostelich, and E. Stone, {\it
Physica D} 96 (1996), 132.

\bibitem{Qin94} F. Qin, E.E. Wolf, and H.-C. Chang, Phys. Rev. Lett.
{\bf 72} (1994), 1459.

\bibitem{PRLprediction} C. L\'{o}pez, A. \'{A}lvarez, and E.
Hern\'{a}ndez-Garc\'{\i}a, Phys. Rev. Lett. 
{\bf 85} (2000), 2300.

\bibitem{GRLprediction}
A. \'{A}lvarez, C. L\'{o}pez, M. Riera, E.
Hern\'{a}ndez-Garc\'{\i}a, J. Tintor\'{e}, Geophys. Res. Lett. 
{\bf 27} (2000), 2709.

\bibitem{Sirovich87}
L. Sirovich and J.D. Rodr\'{\i}guez, Phys. Lett. A {\bf 120}
(1987), 211.

\bibitem{Sirovich89}
L. Sirovich, Physica D {\bf 37} (1989), 126.

\bibitem{Rodriguez90}
J.D. Rodr\'{\i}guez and L. Sirovich, Physica D {\bf 43} (1990), 77.

\bibitem{Deane91}
A.E. Deane, I.G. Kevrekidis, G.E. Karniadakis, and S.A. Orszag,
Phys. Fluids A {\bf 3} (1991), 2337.

\bibitem{Sahan97} R.A. Sahan, A. Liakopoulos, H. Gunes, Phys.
Fluids {\bf 9} (1997), 551.

\bibitem{Aubry88}
N. Aubry, P. Holmes, J.L. Lumley, E. Stone, J. Fluid Mech. {\bf
192} (1988), 115.

\bibitem{Aubry89} N. Aubry, P. Holmes, J.L. Lumley, E. Stone,
Physica D {\bf 37} (1989), 1.

\bibitem{assimilation} M. Ghil, {\it Dynamic Meteorology: Data Assimilation
Methods}, Springer-Verlag, New York, 1981.

\bibitem{Hasselmann}
K. Hasselmann, J. Geophys. Res. {\bf 93} (1988), 11015.

\bibitem{Uhl}
C. Uhl, R. Friedrich, and H. Haken, Phys. Rev. E {\bf 51} (1995),
3890.

\bibitem{Kwasniok97}
F. Kwasniok, Phys. Rev. E {\bf 55} (1997), 5365.

\bibitem{Achatz}
U. Achatz, G. Schmitz, and K.-M. Greisiger, J. Atmos. Sci. {\bf
52} (1995), 3201.

\bibitem{Kwasniok96}
F. Kwasniok, Physica D {\bf 92} (1996), 28.

\bibitem{AVISO} See the different numbers of the AVISO Newsletter,
available on-line at {\tt
http://sirius-ci.cst.cnes.fr:8090/HTML/information/general/welcome.html}.

\bibitem{Ay}
The MSLA products are supplied by the CLS Space Oceanography Division
(France), 
with financial support from the CEO programme (Center for Earth
Observation) and Midi-Pyr\'en\'ees regional council. CD ROMs are
produced by the AVISO/Altimetry operations center. The ERS products
were generated as part of the proposal {\it Joint analysis of
ERS-1, ERS-2 and TOPEX/POSEIDON altimeter data for oceanic
circulation studies } selected in response to the Announcement of
Opportunity for ERS-1/2 by the European Space Agency (Proposal
code: A02.F105);
P. Y. Le Traon, J. Stum, J. Dorandeu, P. Gaspar and P. Vincent,
J. Geophys. Res. {\bf 99} (1994), 24619;
P. Y. Le Traon and F. Ogor, J. Geophys. Res. {\bf 103} (1998), 8045.


\bibitem{tintore} 
U. End, J. Font, G. Kraichman, C. Millot,
M. Rhein and J. Tintor\'e,
Prog. in Oceanog. {\bf 44}  (1999), 37.



\bibitem{Bou1}
C. Bouzinac, J. Vazquez  and J. Font,  J. Geophys. Res. 
{\bf 103} (1998), 8059.

\bibitem{NBCM}
G. R. North, T. L. Bell, R. R. Cahalan and F.J. Moeng,  Mon.
Wea. Rev. {\bf 110} (1982), 699.

\bibitem{Algers}
C. Bouzinac, {\it PhD. Thesis}, U. Pierre et Marie Curie, Paris,
1997.

\bibitem{BK}
F. P. Bretherton and M. Karweit, {\it Mid-ocean mesoscale modeling}
in {\it Modeling of transient or intermediate scale phenomena},
National Academy of Science  (1983), 237.

\bibitem{Ped}
J. Pedlosky, {\it Geophysical Fluid Dynamics (2nd edition)}.
Springer-Verlag, New York, 1987.

\bibitem{Le}
M. Lesieur, {\it Turbulence in Fluids}. Kluwer, Dordretch, 1990.

\bibitem{Benoit}
B. Cushman-Roisin, {\it Introduction to Geophysical Fluid Dynamics}.
Prentice-Hall, New Jersey, 1994.

\bibitem{dstool}
J. Guckenheimer, A. Back, J. Guckenheimer, M. Myers, F. Wicklin
and P. Worfolk, {\it dstool: Computer Assisted Exploration of
Dynamical Systems}, Notices of the American Mathematical Society,
{\bf 39} (1992), 303.

\end{thebibliography}
\end{document}